\documentclass[5p, twocolumn]{elsarticle}

\usepackage{lineno,hyperref}

\journal{Journal of Power Sources}

\usepackage{todonotes}
\usepackage{amsmath}
\usepackage{placeins}
\usepackage{booktabs}
\usepackage{lipsum}
\usepackage{supertabular}
\usepackage{todonotes}
\usepackage{fixltx2e} 
\usepackage{gensymb}
\usepackage{multirow}
\usepackage{afterpage}

\usepackage{color}

\begin{document}

\begin{frontmatter}

\title{On-board monitoring of 2-D spatially-resolved temperatures in cylindrical lithium-ion batteries: Part II. State estimation via impedance-based temperature sensing}

\author{Robert R. Richardson, Shi Zhao and David A. Howey\fnref{myfootnote}}
\address{Department of Engineering Science, University of Oxford, Oxford, UK}
\fntext[myfootnote]{E-mail: \{robert.richardson, shi.zhao, david.howey\} @eng.ox.ac.uk.}

\begin{abstract}
	
	Impedance-based temperature detection (ITD) is a promising approach for rapid estimation of internal cell temperature based on the correlation between temperature and electrochemical impedance.
	Previously, ITD was used as part of an Extended Kalman Filter (EKF) state-estimator in conjunction with a thermal model to enable estimation of the 1-D temperature distribution of a cylindrical lithium-ion battery.
	Here, we extend this method to enable estimation of the 2-D temperature field of a battery with temperature gradients in both the radial and axial directions.
	
	An EKF using a parameterised 2-D spectral-Galerkin model with ITD measurement input (the imaginary part of the impedance at 215 Hz) is shown to accurately predict the core temperature and multiple surface temperatures of a 32113 LiFePO\textsubscript{4} cell, using current excitation profiles based on an Artemis HEV drive cycle. The method is validated experimentally on a cell fitted with a heat sink and asymmetrically cooled via forced air convection.
	
	A novel approach to impedance-temperature calibration is also presented, which uses data from a single drive cycle, rather than measurements at multiple uniform cell temperatures as in previous studies. This greatly reduces the time required for calibration, since it overcomes the need for repeated cell thermal equalization.

\end{abstract}


\end{frontmatter}

\section*{Highlights}
\begin{itemize}
	\item{Demonstration of impedance-based temperature sensing for 2-D problems.}
	\item{Imaginary part of impedance at 215 Hz used as measurement input to EKF state estimator.}
	\item{Low-order 2-D spectral-Galerkin method used for efficient thermal modelling.}
	\item{Experimental validation against internal and surface thermocouple measurements.}
	\item{Impedance-temperature calibration achieved using drive cycle data.}
\end{itemize}

\section{Introduction}

\begin{figure}
	\centering
	\includegraphics[width=1\columnwidth]{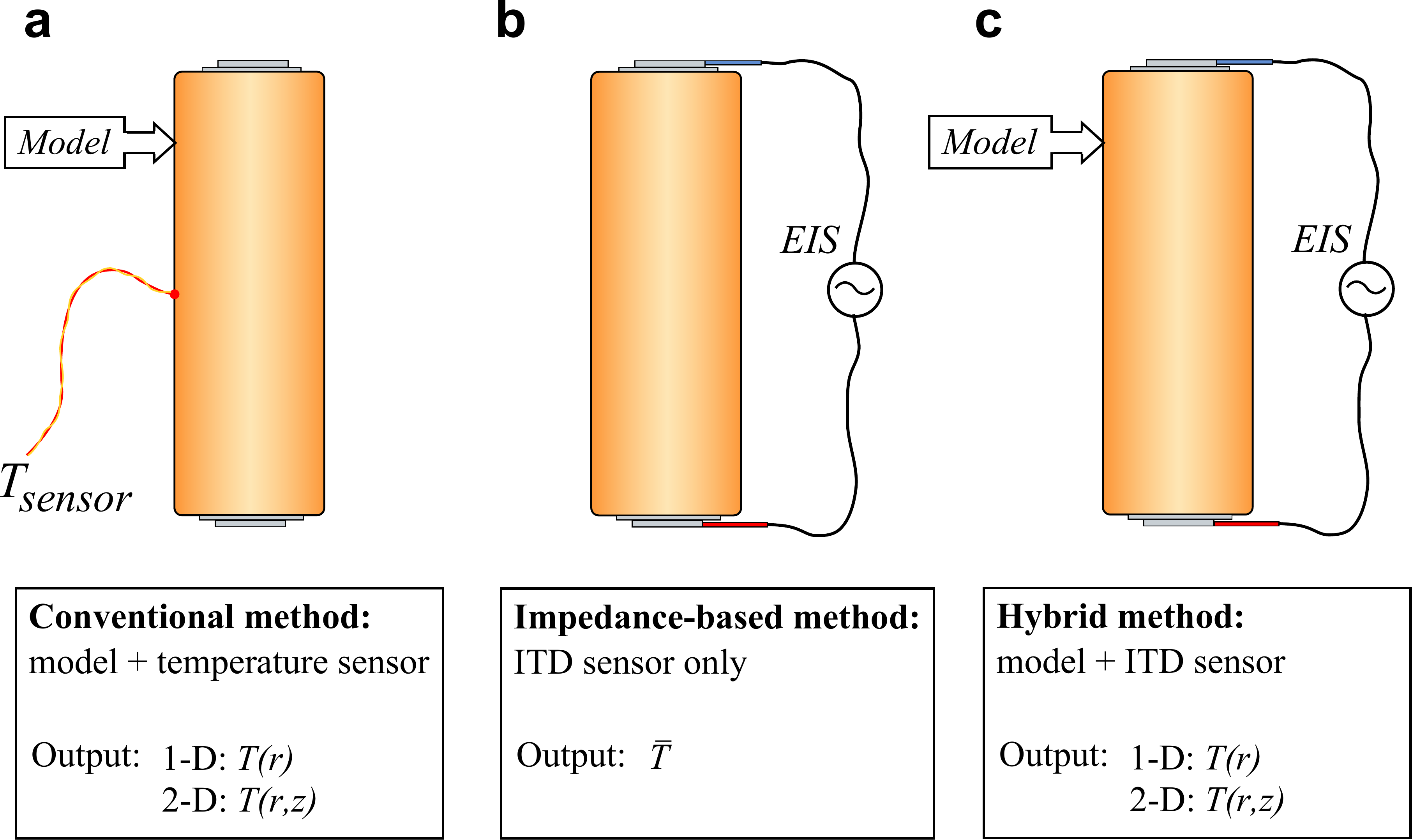}
	\caption{Schematic of approaches to battery temperature monitoring. a) conventional approach, b) impedance-based approach and c) the hybrid approach used in the present study.}
	\label{fig:ITD_schematic}
\end{figure}

Monitoring the temperature of Li-ion batteries during operation is critical for safety and control purposes.
The conventional approach to temperature estimation is to use numerical electrical-thermal models coupled with online measurements of the cell surface temperature and/or the temperature of the heat transfer medium~\cite{Forgez2010a} (Figure~\ref{fig:ITD_schematic})a).
Using this approach in conjunction with state estimation techniques such as Kalman filtering, the cell internal temperature may be estimated with high accuracy~\cite{Kim2013,Kim2014b,Lin2013f,Lin2014}.
However, large battery packs may contain several thousand cells~\cite{Pesaran2009}, and so the requirement for surface temperature sensors on every cell represents substantial instrumentation cost.
As a result, EV manufacturers often use fewer temperature sensors than are required to achieve full observability of the pack~\cite{lin2014temperature}.
Moreover, rapid fluctuations in internal temperature may not be registered by surface mounted temperature sensors, regardless of the sampling frequency. This may mean thermal runaway cannot be detected, since associated timescales are often shorter than those associated with heat conduction through the cell~\cite{santhanagopalan2009analysis}. Consequently, surface mounted temperature sensors even when used with a thermal model may be insufficient to track internal temperature or predict thermal runaway.
One approach to overcome these problems is to embed  flexible thin film micro-temperature sensors within the cell to enable in-situ internal temperature measurement~\cite{lee2011situ,mutyala2014situ,lee2015flexibleA,lee2015flexibleB,martiny2014development}.
Whilst this has some obvious advantages, the additional manufacturing and instrumentation requirements would significantly increase the cost and complexity of the system.

An alternative approach to temperature estimation uses electrochemical impedance spectroscopy (EIS) measurements at one or several frequencies to directly infer the internal cell temperature~\cite{Srinivasan2011c,Srinivasan2012a,Schmidt2013a,Richardson2014,Raijmakers2014d,richardson2015sensorless,Zhu2015,spinner2015expanding,Koch2015} (Figure~\ref{fig:ITD_schematic})b).
This exploits the fact that impedance is a function of the cell internal temperature.
For brevity, we refer to the use of impedance to infer information about the internal temperature as “impedance-temperature detection (ITD)”.
ITD has promise for practical applications, since methods capable of measuring EIS spectra using existing power electronics in a vehicle or other applications have been developed~\cite{Howey2014a}.
Most ITD studies assume the internal temperature is uniform~\cite{Srinivasan2011c,Srinivasan2012a}, or explicitly acknowledge that the impedance measurement alone only predicts the average internal temperature~\cite{Schmidt2013a}.
Our recent work showed that ITD could be used in conjunction with a thermal model in a `hybrid' ITD-based state estimation scheme, to enable estimation of the temperature \emph{distribution} of the cell~\cite{richardson2015sensorless} (Figure~\ref{fig:ITD_schematic}c).
The method is therefore analogous to the conventional state-estimation method, but the measurement input consists of ITD rather than a surface temperature measurement.
However, that work relied on the assumption of 1-D conditions, which are unlikely to arise in all in situations.

In the present paper, we extend the ITD approach to problems involving 2-D thermal dynamics, using the low-order thermal model presented in Part I .

\section{Impedance-based temperature sensing}

The electrochemical impedance $Z(\omega) = Z'(\omega)+ j Z'' (\omega)$ of lithium-ion cells is a function of temperature, state of charge (SoC), and state of health (SoH). Within an appropriate frequency range, however, the dependence on SoC and SoH is negligible and the impedance can thus be used to infer information about the cell temperature. Previous ITD studies have used as a temperature-dependent parameter (TDP) the real part of the impedance at a specific frequency~\cite{richardson2015sensorless,Schmidt2013a}, the imaginary part of the impedance at a specific frequency~\cite{spinner2015expanding}, the phase shift at a specific frequency~\cite{Srinivasan2011c,Srinivasan2012a}, and the intercept frequency~\cite{Raijmakers2014d}.
The issue of which TDP and excitation frequency are most suitable for temperature inference is still an open question, with various arguments in favour of each~\cite{Koch2015,beelen2015improved}.
In the present study, we use the imaginary part of the impedance at f = 215 Hz as the TDP, {since this was found to give superior results to the real part for the studied cell.}
However, {in principle} the method is not limited to this option and could also be applied using any TDP.

The principle of operation of ITD is to relate the cell impedance to the mean%
\footnote{Note that, since the impedance temperature relationship is non-linear (as demonstrated in~\cite{troxler2014effect}), the impedance is not strictly indicative of the volume average temperature, but rather it is related to an EIS-based volume average temperature, as defined in~\cite{richardson2015sensorless}. However, our results~\cite{richardson2015sensorless}, and those of others~\cite{Schmidt2013a} showed that the error introduced by the assumption that the impedance is directly related to $\overline{T}$ is negligible, provided the temperature gradients are not too large. Hence, for simplicity this non-linearity is neglected in this case.} %
cell temperature by a polynomial fit:
\begin{equation}
Z'' = a_1 + a_2 \overline{T} + a_3 \overline{T}^2
\label{eq:f-2D}
\end{equation}
where $a_1$, $a_2$ and $a_3$ are the constant polynomial coefficients.

In our previous work the coefficients of the polynomial fit were identified via offline impedance measurements at multiple uniform temperatures~\cite{richardson2015sensorless}.
However, here we introduce an improved calibration technique, which uses data from a single drive cycle (see Section~\ref{sec:impedance_cal_1}). This greatly reduces the time required for calibration, since it overcomes the need for cell thermal equalization at multiple temperatures.

Finally, the principle of operation of the hybrid ITD-based state-estimation method is to use ITD as the measurement input to a state-estimator in conjunction with the 2-D thermal model presented in the Part I of this paper \cite{richardson2016on}.

\section{Overview of hybrid method\label{sec:approach_SG2}}

\begin{figure*}[h]
	\centering
	\includegraphics[width=0.75\textwidth]{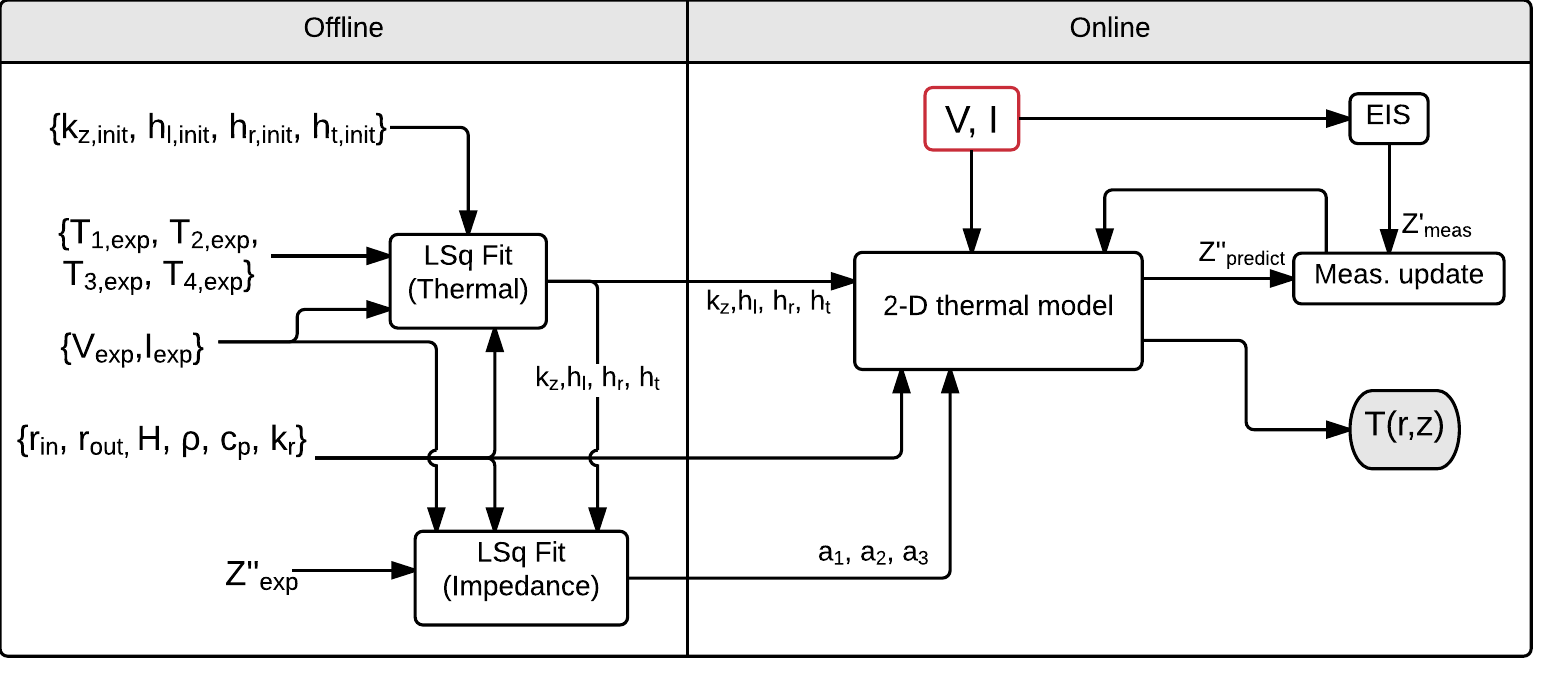}
	\caption{Process flowchart of the ITD-based state estimation method.}
	\label{fig:ITD_TM_process_2D}
\end{figure*}

An overview of the process is shown in Figure~\ref{fig:ITD_TM_process_2D} and described below.

\emph{Offline:}
{The cell dimensions (jelly-roll inner radius, $r_{in}$; outer radius, $r_{out}$; and height, $H$) are measured and therefore known a-priori; the thermal properties (density, $\rho$; specific heat capacity, $c_p$; and radial conductivity, $k_r$) are available from the literature~\cite{fleckenstein2013thermal}, and therefore known a-priori. The remaining thermal properties (axial conductivity, $k_z$; and the convection coefficients on the left end, $h_l$, right end, $h_r$, and curved surface, $h_t$) are then estimated using a recursive least squares fitting algorithm (see Section~\ref{sec:SG2_param}).}
Specifically, the thermal model is simulated using experimental current and voltage data ($V_{exp}$, $I_{exp}$), from a parameterization drive cycle and the temperatures at four locations ($T_1$, $T_2$, $T_3$, and $T_4$) are compared with their corresponding thermocouple measurements. The parameterised model is then used to fit the impedance coefficients (see Section~\ref{sec:impedance_cal_1}). Specifically, the predicted mean temperature ($\overline{T}$) from the model is paired with the corresponding measured impedance value ($Z''$) at each sample. A second order polynomial fit is then applied to the resulting impedance-temperature data, thus identifying coefficients ($a_1$, $a_2$, $a_3$). Finally, the known and identified thermal parameters and the identified impedance coefficients are provided to the online thermal model.

\emph{Online:} The thermal model (Section~\ref{sec:model}) uses online measurements of the voltage and current ($V$, $I$) to predict the heat generation, 2-D temperature distribution, and cell  imaginary impedance at each time step.
This is compared to the measured imaginary impedance which is used to update the state estimate of the model via an Extended Kalman Filter algorithm (EKF) (Section~\ref{sec:SG2_state_est}).

\section{Theory}

\subsection{Problem definition}

\begin{figure}[hbt]
	\centering
	\includegraphics[width=0.8\columnwidth]{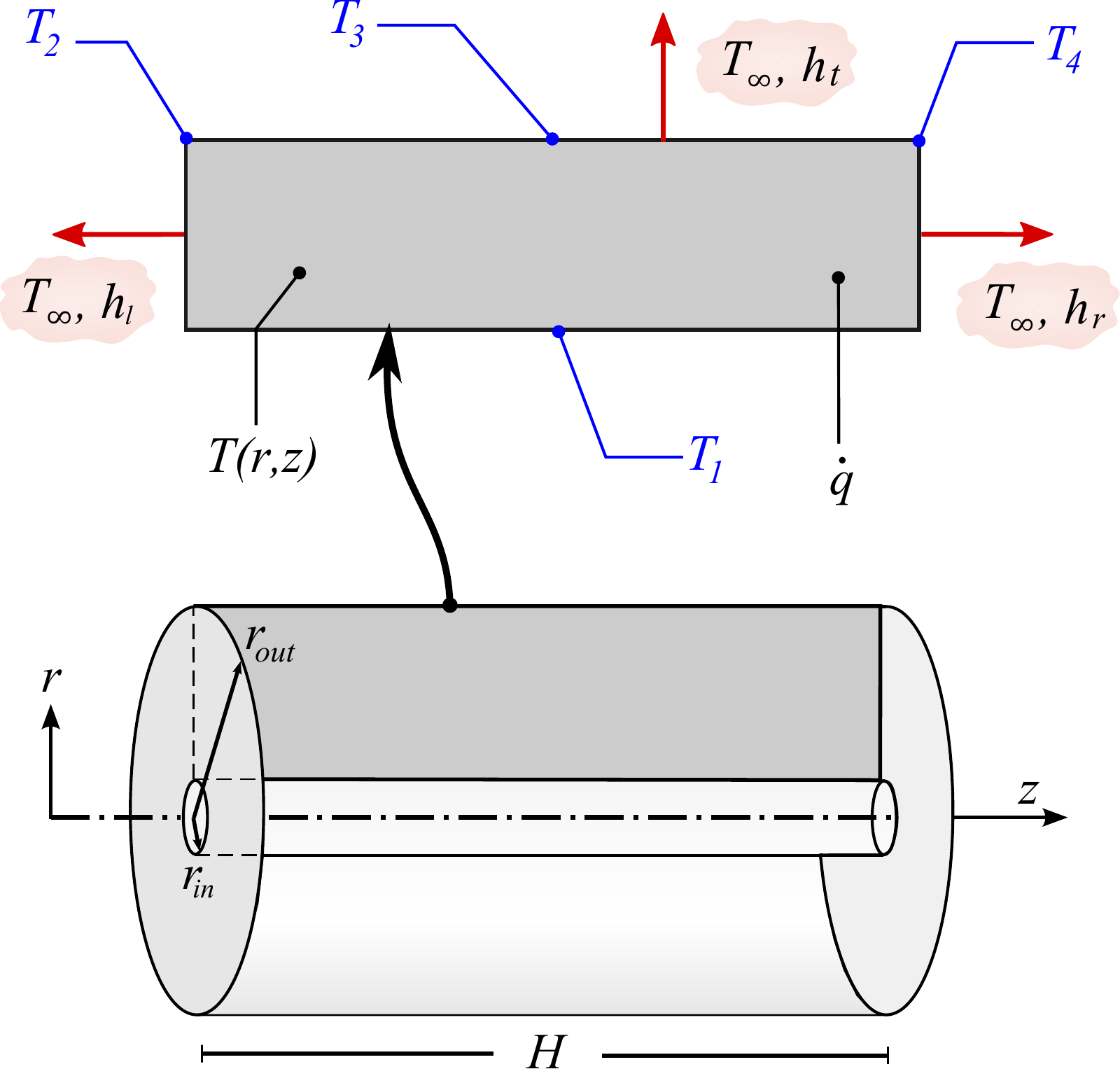}
	\par
	\caption{Schematic of cylindrical cell geometry for the thermal model, showing (\textit{red}) different convection coefficients at each surface and (\textit{blue}) locations of model outputs corresponding to thermocouple sensor locations.}
	\label{fig:schematic}
\end{figure}

The model consists of the transient 2-D energy conservation equation in cylindrical coordinates. Heat generation is assumed to be uniform in space but time-dependant. The multi-layer structure of the battery is treated as a homogeneous solid with anisotropic thermal conductivity in the radial and axial directions. The temperature variation in the azimuthal ($\varphi$) direction is neglected. Convective heat transfer is assumed to occur at the outside surfaces, and there is zero convection at the inner annulus of the jelly roll (i.e. at the central mandrel).
The convection coefficient of the ambient air may be different at each surface, although its free-stream temperature is assumed to be uniform throughout the chamber.
A schematic of the model is shown in Figure~\ref{fig:schematic}.

The model is governed by the following 2-D boundary value problem \cite{hahn2012heat}:
\begin{equation}
\rho c_{p}\frac{\partial T}{\partial t} - k_{r}\frac{\partial^{2}T}{\partial r^{2}} -
\frac{k_{r}}{r}\frac{\partial T}{\partial r} - k_{z}\frac{\partial^{2}T}{\partial z^{2}} = q
\label{eq:original_eq}
\end{equation}
where $t$ is time and $r$ and $z$ are the position coordinates in the radial and axial directions respectively. The functions $T(r,z,t)$ and $q(t)$ are the temperature distribution and volumetric heat generation rate, respectively. The parameters $\rho$ and $c_p$ are the density and specific heat capacity respectively, and $k_r$ and $k_z$ are the anisotropic thermal conductivities in the $r$ and $z$ directions. 
The boundary conditions are given by:
\begin{subequations}
	\begin{align}	
	\frac{\partial T}{\partial r} & =  -\frac{h_{r}}{k_{r}}(T-T_{\infty})
	\text{\quad at $r = r_{out}$} \label{eq:BC1_chap7} \\
	\frac{\partial T}{\partial r} & =  0
	\text{\quad at $r = r_{in}$} \\
	\frac{\partial T}{\partial z} & = 	-\frac{h_{t}}{k_{z}}(T-T_{\infty})
	\text{\quad at $z = H$} \label{eq:BC3}\\
	\frac{\partial T}{\partial z} & = 	\frac{h_{b}}{k_{z}}(T-T_{\infty})
	\text{\quad at $z = 0$} \label{eq:BC4}
	\end{align}
\end{subequations}
where $T_{\infty}$ is the free-stream air temperature of the chamber, and $\{h_{\sigma}; \, \sigma = t, r \text{ and } l \}$ are the convection coefficients at the top, right and left surfaces.
Note that the convection coefficient at the bottom surface is set to zero (since this corresponds to the inner radius of the cell jelly roll, which is not exposed to cooling). $h_t$ corresponds to the curved surface of the cell, whilst $h_l$ and $h_r$ correspond to left and right ends of the cell. Note that the placement of the heat sink (see later) results in an increased value of $h_l$.

As in our previous work~\cite{richardson2015sensorless}, we consider only ohmic heat generation, given by:
\begin{align}
Q=I(V-U_{OCV}).
\label{eq:Q}
\end{align}
The entropic heat is neglected because (i) the net reversible heat would be close to zero when the cell is operating in HEV mode and (ii) the HEV drive cycles employed in this study operate the cell within a small range of SoC ($47 - 63 \, \%$) and hence the entropic heat is small~\cite{Forgez2010a}.
The heat generation is assumed to be uniformly distributed throughout the cell volume, hence the volumetric heat generation is given by:
\begin{equation}
q = \frac{Q}{V_b},
\end{equation}
where $V_b$ is the cell volume.

\subsection{Spectral-Galerkin model\label{sec:model}}

The above problem can be simulated using the spectral-Galerkin approach described in Part I of this paper. The model can be expressed in state space form:
\begin{align}
\mathbf{E}\mathbf{\dot{x}} = \mathbf{A}\mathbf{x} + \mathbf{B}\mathbf{u}
\label{eq:state-space-exp}
\\
\mathbf{y} = \mathbf{C}\mathbf{x} + \mathbf{T_e}
\end{align}
where the states and the state matrices are as defined in Part I, and the model input is $\mathbf{u} = \left[q(t), \, 1\right]^T$.
Four temperature outputs are chosen corresponding to the positions $T_{1} = T(r=r_{in},z=H/2)$, $T_{2} = T(r=r_{out},z=0)$, $T_{3} = T(r=r_{out},z=H/2)$ and $T_{4} = T(r=r_{out},z=H)$ (see Figure~\ref{fig:schematic}).
These outputs were chosen to match the thermocouple measurements in the experimental setup as described in the following section.

Lastly, the cell imaginary impedance at the selected frequency is also calculated as an output. This is obtained from the mean temperature using eq.~\ref{eq:f-2D}. $\overline{T}$ is itself a function of the model states, $\mathbf{x}$ as described in the companion paper. Hence an expression for the impedance as a function of the cell states is obtained,
\begin{equation}
	Z'' = f(\mathbf{x}).
	\label{eq:f_x}
\end{equation}
The computed impedance is used as the measurement input in the state estimation algorithm described next.

\subsection{State estimation\label{sec:SG2_state_est}}
The state estimation consists of an extended Kalman filter (EKF), for estimating the temperatures at each of the four thermocouple locations, with the cell impedance as measurement input. The EKF is equivalent to that employed in \cite{richardson2015sensorless} for the 1-D case, except that here it is applied to the 2-D SG model.

We firstly modify eq.~(\ref{eq:state-space-exp}) by rewriting it as an explicit state-space model in discrete time:
\begin{equation}
\mathbf{x}_{k+1} =\bar{\mathbf{A}}\mathbf{x}_{k}+\bar{\mathbf{B}}\mathbf{u}_{k}+\mathbf{v}_{k},
\end{equation}
where
$\bar{\mathbf{A}}$ and $\bar{\mathbf{B}}$ are system matrices in the discrete-time domain, given by
\begin{align}
\bar{\mathbf{A}} & = e^{\left((\mathbf{E}^{-1}\mathbf{A})\Delta t\right)},\\
\bar{\mathbf{B}} & = (\mathbf{E}^{-1}\mathbf{A})^{-1}(\bar{\mathbf{A}}-\mathbf{I})(\mathbf{E}^{-1}\mathbf{B}),
\end{align}
where $\Delta t$ is the sampling time of 1 s.
We then set the impedance as the model output
\begin{equation}
y_{k} = f(\mathbf{x}_{k})+n_{k},
\label{eq:SG2_y}
\end{equation}
where $f(\mathbf{x}_{k})$ is the non-linear function relating the state vector to the impedance measurement (i.e. eq.~\ref{eq:f_x}), and $\mathbf{v}_{k}$ and $n_{k}$ are the process and measurement noise, respectively. Their corresponding covariance matrices are $\mathbf{R}^{\mathbf{v}}$ and $R^{n}$.
Note that, although the impedance is the model output for the purpose of the EKF algorithm, the temperatures at each of the four thermocouple locations are also computed from the identified states at each time step, for validation against the thermocouple measurements.

The time update processes are then given by:
\begin{align}
\hat{\mathbf{x}}_{k}^{-} & =\bar{\mathbf{A}}\hat{\mathbf{x}}_{k-1}+\bar{\mathbf{B}}\mathbf{u}_{k-1},\\
(\mathbf{P}_{k}^{\mathbf{x}})^{-} & =\bar{\mathbf{A}}\mathbf{P}_{k-1}^{\mathbf{x}}\bar{\mathbf{A}}^{T}+\mathbf{R}^{\mathbf{v}},
\end{align}
where $\mathbf{\hat{x}}_{k}^{-}$ and $\mathbf{\hat{x}}_{k}$ are the \emph{a priori }and \emph{a posteriori} estimates of the state, and $(\mathbf{P}_{k}^{\mathbf{x}})^{-}$ and $\mathbf{P}_{k-1}^{\mathbf{x}}$ are the corresponding\emph{ }error covariances.
Since the relationship between impedance and the cell state is non-linear, the measurement model must be linearised about the predicted observation at each measurement. The measurement update equations are:
\begin{align}
\mathbf{K}_{k}^{\mathbf{x}} & =(\mathbf{P}_{k}^{\mathbf{x}})^{-}(\mathbf{H}_{k}^{\mathbf{x}})^{T}\left(\mathbf{H}_{k}^{\mathbf{x}}(\mathbf{P}_{k}^{\mathbf{x}})^{-}\mathbf{(H}_{k}^{\mathbf{x}})^{T}+R^{n}\right)^{-1},\\
\hat{\mathbf{x}}_{k} & =\hat{\mathbf{x}}_{k}^{-}+\mathbf{K}_{k}^{\mathbf{x}}\left(z_{k}-f(\mathbf{\hat{x}}_{k}^{-})\right),\label{eq:SG2_meas_up_2}\\
\mathbf{P}_{k}^{\mathbf{x}} & =(\mathbf{I}-\mathbf{K}_{k}^{\mathbf{x}}\mathbf{H}_{k}^{\mathbf{x}})(\mathbf{P}_{k}^{\mathbf{x}})^{-},
\end{align}
where $\mathbf{K}_{k}^{\mathbf{x}}$ is the Kalman gain
for the state,\emph{ }and $\mathbf{H}_{k}^{\mathbf{x}}$ is the Jacobian
matrix of partial derivatives of $f$ with respect to $\mathbf{x}$:
\begin{align}
\mathbf{H}_{k}^{\mathbf{x}}=\frac{\partial f(\mathbf{x}_k)}{\partial\mathbf{x}_k}\Bigg|_{\mathbf{x}_k=\hat{\mathbf{x}}_{k}^{-}}.
\label{eq:SG2_jacob}
\end{align}
The above algorithm can be simplified to a standard Kalman Filter (KF) by omitting the linearisation step (eq.~\ref{eq:SG2_jacob}) and replacing $f(\cdot)$ in eqs.~\ref{eq:SG2_y} and \ref{eq:SG2_meas_up_2} with an appropriate linear operator. Such a KF is applied later using the surface temperature, $T_3$, as measurement input for comparison with the EKF using $Z''$.

\section{Experimental\label{sec:SG2_exp}}

\subsection{Setup}

\begin{figure*}[h]
	\centering
	\includegraphics[width=0.9\textwidth]{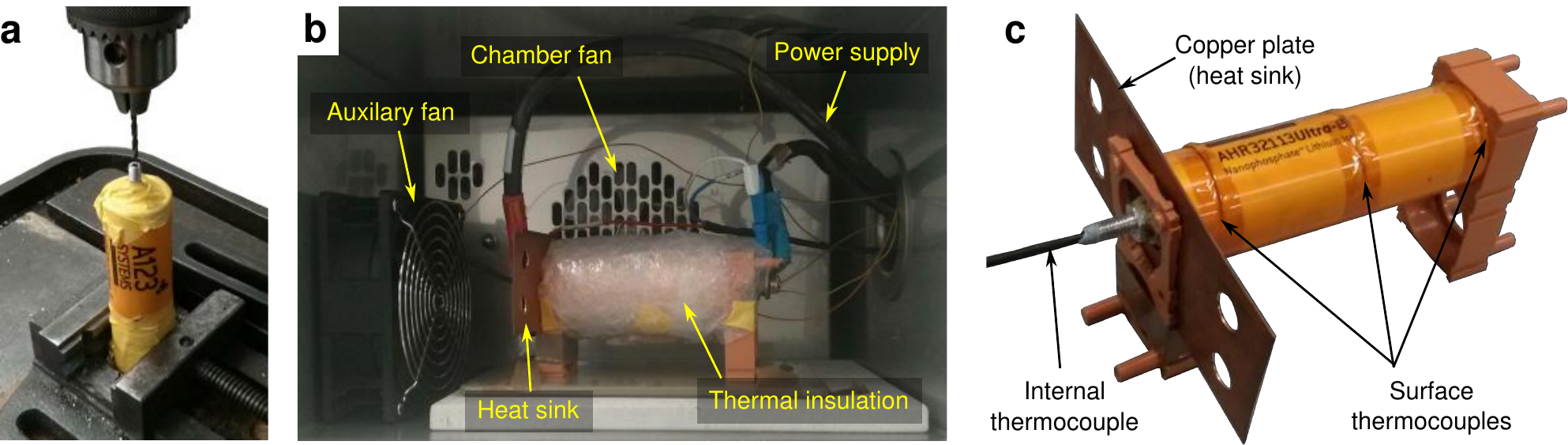}
	\par \vspace{0.9 cm}
	\includegraphics[width=0.9\textwidth]{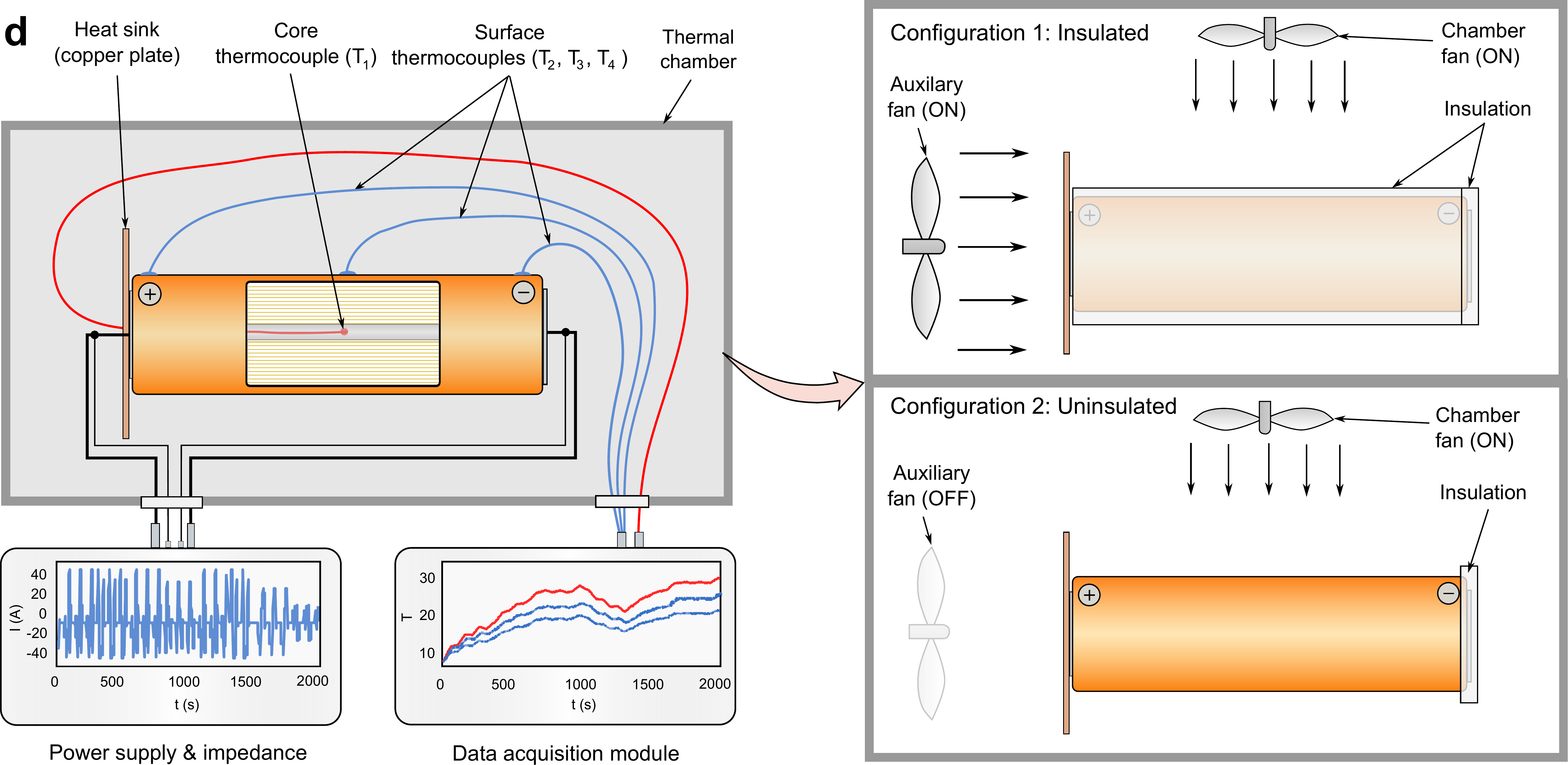}
	\caption{Experimental setup: (a) cell drilling procedure, (b) insulated cell (Config.\ 2) inside thermal chamber, (c) uninsulated cell with heat sink showing thermocouple locations, (d) schematic diagram showing (\emph{left}) cell in chamber with cutaway view showing core and jelly roll; (\emph{right}) the two cooling configurations.}
	\label{fig:experimental_setup_2D}
\end{figure*}

Experiments were carried out with a 4.4~Ah cylindrical cell (A123 Model AHR 32113 Ultra-B) with LiFePO\textsubscript{4} positive electrode and a graphite negative electrode.
A large form factor cell was used in order to ensure measurable 2-D effects. The properties of the cell are given in Table~\ref{table:cell-properties-2D}. The cell was fitted with four thermocouples, three on the surface and another inserted into the core via a hole which was drilled in the positive electrode end.
The thermocouple locations correspond to the model output locations described previously. Cell cycling and impedance measurements were carried out using a Biologic HCP-1005 potentiostat/booster. The impedance was measured using Galvanostatic Impedance Spectroscopy (GEIS) with a 200 mA peak-to-peak perturbation current. The environmental temperature was controlled with a Votsch VT4002 thermal chamber. The chamber includes a fan which runs continuously at a fixed speed during operation. Photos of the test equipment and a schematic of the experimental setup are shown in Figure~\ref{fig:experimental_setup_2D}.

Two different cooling configurations were tested (see Figure~\ref{fig:experimental_setup_2D}d). In Config.\ 1 the right end and entire curved surface of the cell are thermally insulated, and a heat sink fixed to the left end. An additional (auxiliary) fan is placed inside the chamber as shown in Figure~\ref{fig:experimental_setup_2D}b. In this case, both the built-in chamber fan and the auxiliary fan run continuously throughout the duration of the experiments. This setup is designed to achieve maximum cooling from the left end of the cell (heat sink)  whilst minimising radial heat transfer.
In Config.\ 2, only the right end of the cell is insulated, and the auxiliary fan is switched off. This setup allows for greater radial heat flux.

\begin{table}[h]
	\centering
	\caption{Properties of the lithium-ion cell used for validation.}
	\begin{tabular}{l l}
		\toprule
		Model & A123 AHR-32113 \\
		Anode material & Graphite \\
		Cathode material & LiFePO\textsubscript{4} \\
		Nominal voltage  & 3.3 V \\
		 Nominal capacity & 4.4 Ah \\
		 Jelly-roll length ($H$) & 100 mm \\
		 Jelly-roll outer radius ($r_{out}$) & 16 mm \\
		 Jelly-roll inner radius ($r_{in}$) & 1 mm\\
		\bottomrule
	\end{tabular}
	\label{table:cell-properties-2D}
\end{table}

\subsection{Procedure: Full spectrum impedance}
The procedure described here applies to the results presented in Section 6.1.
In order to verify that the impedance at the selected frequency was independent of SoC, full spectrum impedance measurements were carried out at a uniform cell temperature of 10\degree C, over a range of 10-90\% SoC in intervals of 20\%%
\footnote{The setup of Config.\ 2 was used for the full spectrum impedance measurements, although note that the choice between Config.\ 1 and 2 for this step is somewhat arbitrary since the chamber and cell temperatures are allowed to equilibrate before each measurement.}.
The EIS frequency range spanned 5 kHz to 0.1 Hz with 10 frequencies per decade and 3x averaging. The cell capacities at each temperature were first determined with a constant current constant voltage (CCCV) charge/discharge. The SoC was then adjusted to the required values by drawing a 1 C current. At each SoC, the cell was allowed to rest to ensure its temperature had equilibrated, which typically occurred in less than 2 hrs. This was verified when the temperatures registered by the internal and surface thermocouples were within 0.2 \degree C and repeated impedance measurements over a 20 minute period yielded identical results.

\subsection{Procedure: validation experiments\label{sec:SG2_exp_val}}

The procedure described here applies to the results presented in Sections 6.2 - 6.4.
Dynamic experiments were conducted using three different current excitation profiles, denoted HEV-I, HEV-II, and HEV-III. Each cycle was generated by looping over different portions of an Artemis HEV drive cycle, and scaling the applied currents to the range $\pm50$ A.
HEV-I was used to parameterise the thermal model, and HEV-II and HEV-III were used to validate the identified parameters and to demonstrate the temperature estimation technique (see Section~\ref{sec:SG2_param}).

The procedure for the experiments was as follows: impedance measurements at 215 Hz were carried out every 24 s with 4 s pauses before each impedance measurement, and the four thermocouple temperatures were also monitored.
Before each experiment, the SoC was adjusted to 50\% by drawing a 1 C current.
{The temperature of the thermal chamber was set to 8 $^{\circ}$C in order to maintain the cell within a typical operating range (cell temperatures of $\sim$20 to 30\degree C are generally optimal \cite{bandhauer2011critical}).}
The cell was allowed to rest until its temperature equilibrated before each test began.

\section{Results and discussion}

\subsection{Full spectrum impedance\label{sec:SG2_full}}

{Our earlier work demonstrated independence of the impedance at 215 Hz with respect to SoC for a 26650 LiFePO\textsubscript{4} cell (A123 ANR26650 m1-A) over a range of temperatures from -20 to 45\degree C. Since the cell used in the present study is from the same manufacturer and has the same chemistry, it was deemed sufficient to verify SoC independence at a single temperature close to the middle of this range.
Hence, the full spectrum impedance tests were carried out at a uniform cell temperature of 10\degree C (over a range of 10-90\% SoC in intervals of 20\%). The results are provided in the supplementary material. They confirm that the impedance at 215 Hz is approximately independent of SoC, as required.}


\subsection{Parameterisation and validation\label{sec:SG2_param}}

The full set of thermal parameters include: $\rho$, $c_p$, $k_r$, $k_z$ and the three convection coefficients  (one on each end, and one on the cell curved surface), denoted by $h_l$, $h_r$ and $h_t$ (see Fig.~\ref{fig:schematic}). However, three of these parameters were known a priori from the literature:
Fleckenstein et al.~\cite{fleckenstein2013thermal} identified values for the density, $\rho$, the specific heat capacity, $c_p$, and the radial thermal conductivity, $k_r$, for an identical cell using thermal impedance spectroscopy, and so these values are used in the present case.
Hence, parameterisation is only required to estimate the remaining four parameter values: the axial thermal conductivity, $k_{z}$, and the three convection coefficients: $h_l$, $h_r$ and $h_t$.

The measurements from HEV-I (comprising cell current, voltage, and surface and core temperatures, and the chamber temperature) were used for the parameter estimation for both cooling configurations.
{HEV-I was deliberately chosen for the parameterization since it results in slightly higher cell temperatures than HEV-II or HEV-III, and hence ensures that the polynomial fit is applied using the largest temperature range possible.} For Config.\ 1, all four parameters were identified. For Config.\ 2, the axial thermal conductivity, $k_z$, was assumed known a-priori, using the identified value from Config.\ 1 (since $k_z$ is the same in each case), and hence it was only necessary to identify the three convection coefficient parameters.

The parameterisation was carried out using \emph{fmincon} in Matlab to minimise the magnitude of the Euclidean distance between the measured and estimated temperatures for each of the four thermocouples. Concretely, the error between the measured (subscript `exp') and model predictions (subscript `m') at each time step, $k$, is given by,
\begin{multline}
\epsilon(k, \theta) = [T_{1, m}(k, \theta), T_{2, m}(k, \theta), T_{3, m}(k, \theta), T_{4, m}(k, \theta)]
\\
- [T_{1, exp}, T_{2, exp}, T_{3, exp}, T_{4, exp}]
,
\end{multline}
and the parameters are identified by,
\begin{equation}
\theta^* = \arg \min\limits_{\theta} \sum\limits_{k=1}^{N_f} \| \epsilon(k,\theta) \|_2
.
\end{equation}
where $N_f$ is the number of time steps in the cycle.

Table~\ref{tab:Thermal-parameters-2D} presents the thermal parameters, including those known a-priori and those identified via parameterisation.
The convection coefficient values are within the range expected of forced convection air cooling~\cite{Incropera2007a}. The left coefficient is greatest in both cases as expected due to the presence of the copper heat sink. The values of $h_r$ and $h_l$ are greater in Config.\ 1, due to the presence of the auxiliary fan, whereas the value of $h_t$ is greater in Config.\ 2 since the curved surface is uninsulated. The value of $k_z$ is $\sim 55$ times greater than $k_r$; this is typical of cylindrical cells with wound jelly-roll constructions~\cite{fleckenstein2013thermal}.
Uncertainty in the parameter estimation may be attributed to manufacturing variability, error in the heat generation calculation (due to the omission of entropic heating), heat generation in the contact resistances between the cell and connecting wires and/or measurement uncertainty in the temperature.

\begin{table}[h]
	\caption[Thermal parameters for the two cooling configurations]{Thermal parameters for Configs. 1 and 2, including those known a-priori from the literature~\cite{fleckenstein2013thermal} and those identified using the parameterisation cycle.}
	\centering
	\label{tab:Thermal-parameters-2D}
	\begin{tabular}{l l l l l l}
		\toprule
		\multirow{2}{*}{Param.} & \multirow{2}{*}{Units} & \multicolumn{2}{c}{Config.\ 1} & \multicolumn{2}{c}{Config.\ 2} \\
		\cmidrule{3-6}
		& & Known & Id. & Known & Id.   \\ 
		\midrule
		$\rho$ & kg m\textsuperscript{-3} & 2680 & & 2680 & \\
		$c_p$ & J kg\textsuperscript{-1} K\textsuperscript{-1} & 958 & & 958 & \\
		$k_r$ & W m\textsuperscript{-1} K\textsuperscript{-1} & 0.35 &  & 0.35 & \\
		$k_z$ & W m\textsuperscript{-1} K\textsuperscript{-1} & & 19.3 & 19.3 & \\
		$h_l$ & W m\textsuperscript{-2} & & 155 &  & 98.2\\
		$h_r$ & W m\textsuperscript{-2} & & 23.3 &  & 7.2\\
		$h_t$ & W m\textsuperscript{-2} & & 16.9 &  & 56.2\\
		\bottomrule
	\end{tabular}
\end{table}

\begin{figure}[h]
	\centering
	\includegraphics[width=1\columnwidth]{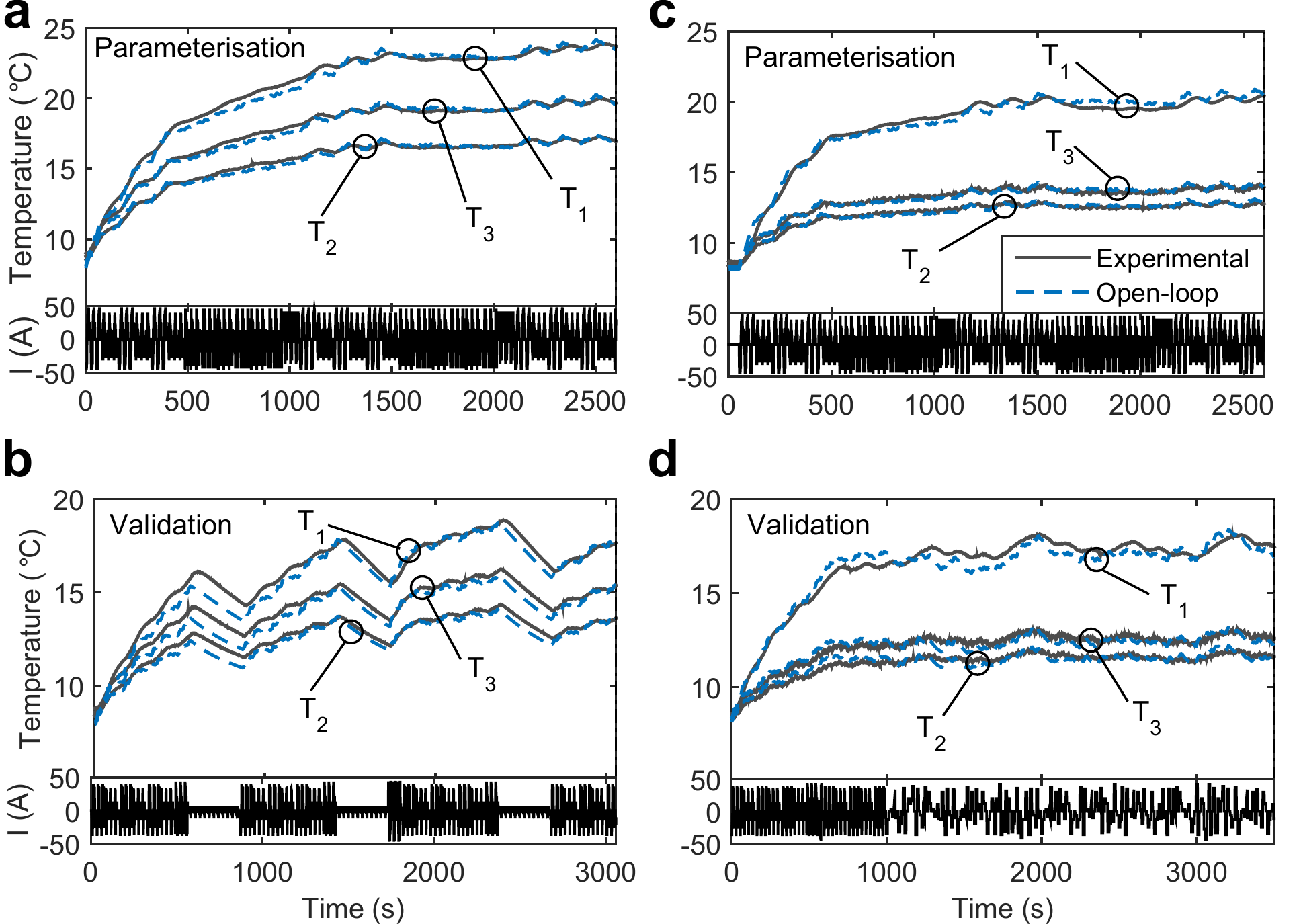}
	\caption[Model parameterisation and validation results]{Comparison between measured and predicted temperatures at the thermocouple locations for the parameterisation and validation cycles. a) Config.\ 1, parameterisation with HEV-I; b) Config.\ 1, validation with HEV-II; c) Config.\ 2, parameterisation with HEV-I; d) Config.\ 2, validation with HEV-III. For clarity, $T_4$ is omitted from these and subsequent plots since it is very close in value to $T_3$.}
	\label{fig:param_validation_2d}
\end{figure}

The measured core and surface temperatures (subscript `exp') and the corresponding model predictions (subscript `m') for the parameterised model for Config.\ 1 are shown in Fig.~\ref{fig:param_validation_2d}a. The model with identified parameters was validated against the second current excitation profile, HEV-II (Fig.~\ref{fig:param_validation_2d}b).
For Config.\ 2, HEV-I was used as the parameterisation cycle and HEV-III for validation; these results are shown in Figures~\ref{fig:param_validation_2d}c and \ref{fig:param_validation_2d}d respectively.
Note that, for clarity, $T_4$ is omitted from these and subsequent plots since it is very close in value to $T_3$.
The root-mean-square errors (RMSE) in each case are shown in table~\ref{tab:param_validation_2d_table}.
The errors in the validation tests are only marginally greater than those in the parameterisation tests, indicating that the estimation is satisfactory.

\begin{table}[h]
	\caption[RMS errors in the parameterisation and validation cycles]{RMS errors in the parameterisation and validation cycles for the two cooling configurations. Units: \degree C.}
	\centering
	\label{tab:param_validation_2d_table}
	\begin{tabular}{l l l l l }
		\toprule
		\multirow{2}{*}{Sensor} & \multicolumn{2}{c}{Config.\ 1} & \multicolumn{2}{c}{Config.\ 2} \\
		\cmidrule{2-5}
		& Param. & Validation & Param. & Validation   \\ 
		\midrule
		$T_1$ & 0.353 & 0.659 & 0.384 & 0.458 \\
		$T_2$ & 0.255 & 0.448 & 0.213 & 0.282 \\
		$T_3$ & 0.245 & 0.434 & 0.209 & 0.279 \\
		$T_4$ & 0.188 & 0.310 & 0.176 & 0.234 \\
		\bottomrule
	\end{tabular}
\end{table}

\subsection{Impedance calibration\label{sec:impedance_cal_1}}

\begin{figure}[h]
	\centering
	\includegraphics[width=0.86\columnwidth]{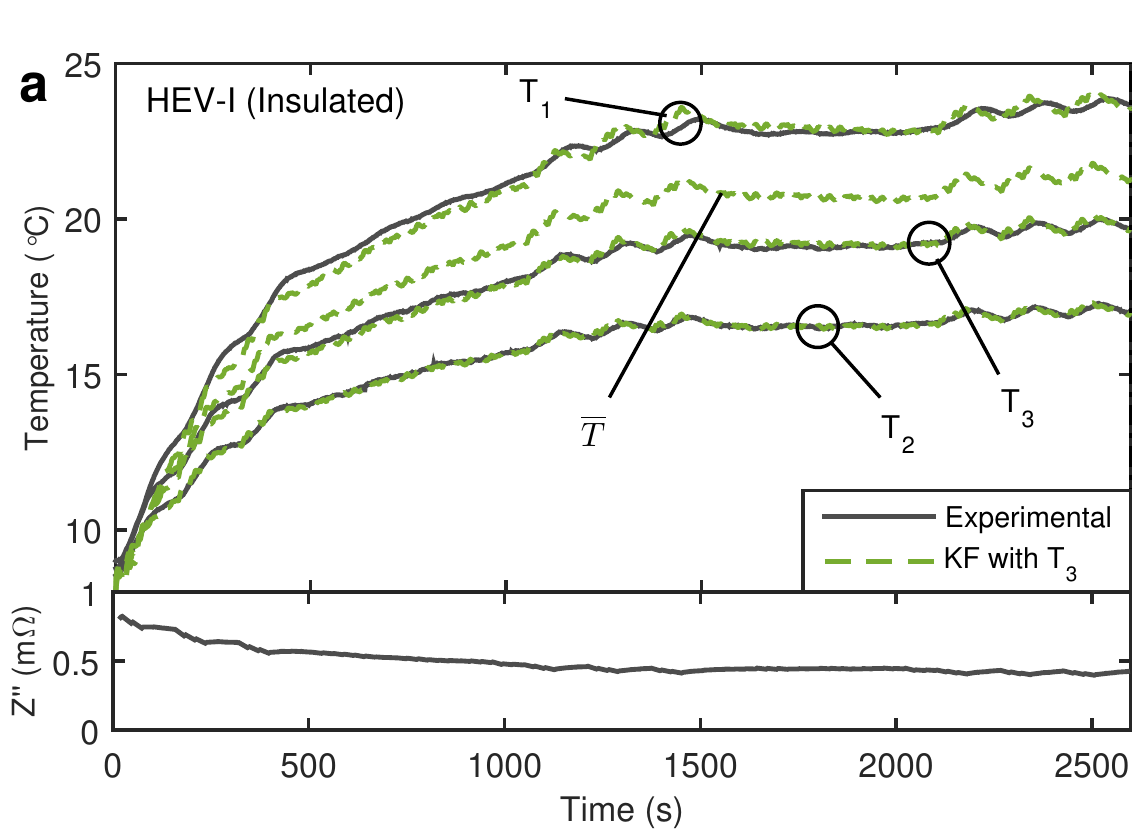}
	\includegraphics[width=0.86\columnwidth]{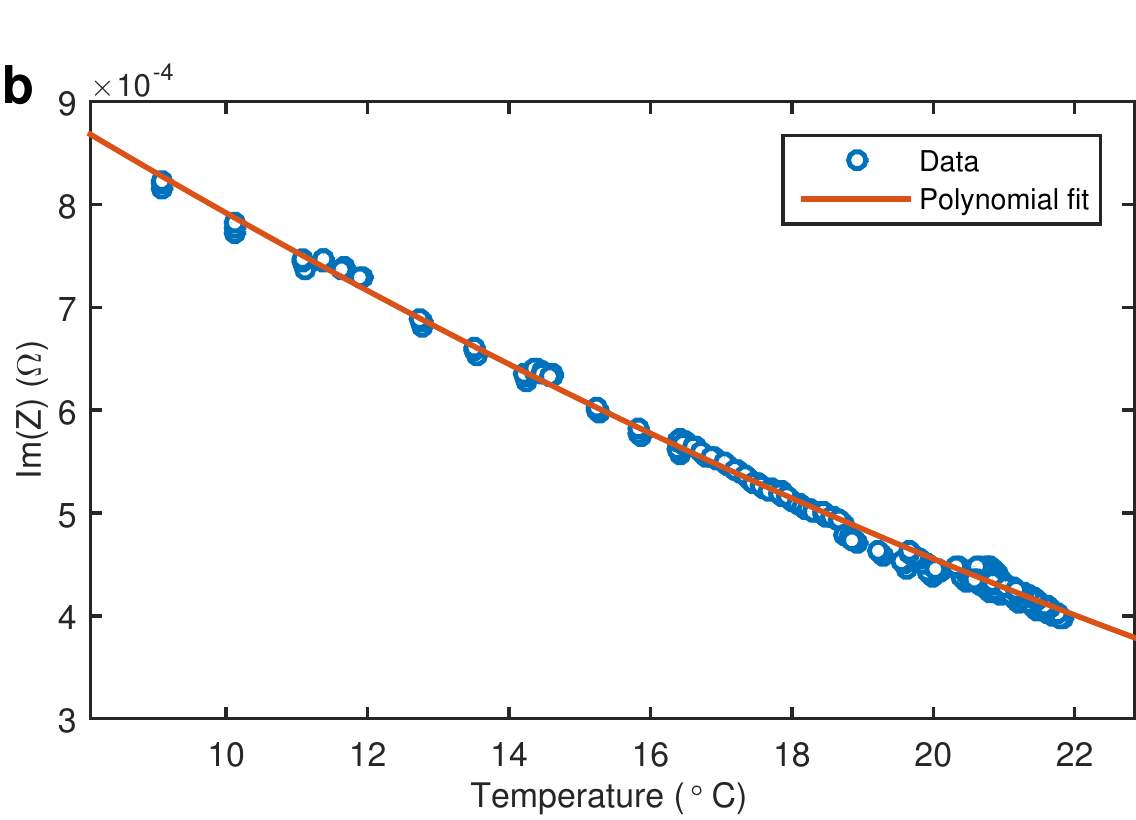}
	\caption{Identification of impedance-temperature polynomial fit. a) Parameterisation drive cycle results;  b) Resulting polynomial fit of $Z''$ against $\overline{T}$.}
	\label{fig:polyfit_exp_insulated}
\end{figure}

The calibration of the impedance-temperature polynomial coefficients ($a_1$, $a_2$ and $a_3$) is achieved as follows. The current/voltage data from Config.\ 1, HEV-I is applied to the parameterised model  (Fig.~\ref{fig:polyfit_exp_insulated}a. A KF is applied to the model using the surface temperature ($T_3$) as measurement input\footnote{Note that the open-loop model could alternatively have been used here and is sufficient for the purpose of impedance calibration; however, the KF was applied since it further increases the model accuracy.}. The predicted $\overline{T}_m$ output from the KF at each measurement step is paired with the measured $Z_{exp}''$ value from the HEV-I experimental data. A second order polynomial fit is then applied to the {$Z_{exp}''$ vs.\ $\overline{T}_m$} data (Fig.~\ref{fig:polyfit_exp_insulated}b). Fig.~\ref{fig:polyfit_exp_insulated} shows that a second order fit is capable of closely approximating the measured data. It should be noted that an Arrhenius fit could also be obtained but the polynomial fit is sufficiently accurate and facilitates faster online computation.

\subsection{State estimation\label{sec:SG2_t1}}

\afterpage{
	\begin{figure*}
		\centering
		\includegraphics[width=0.89\textwidth]{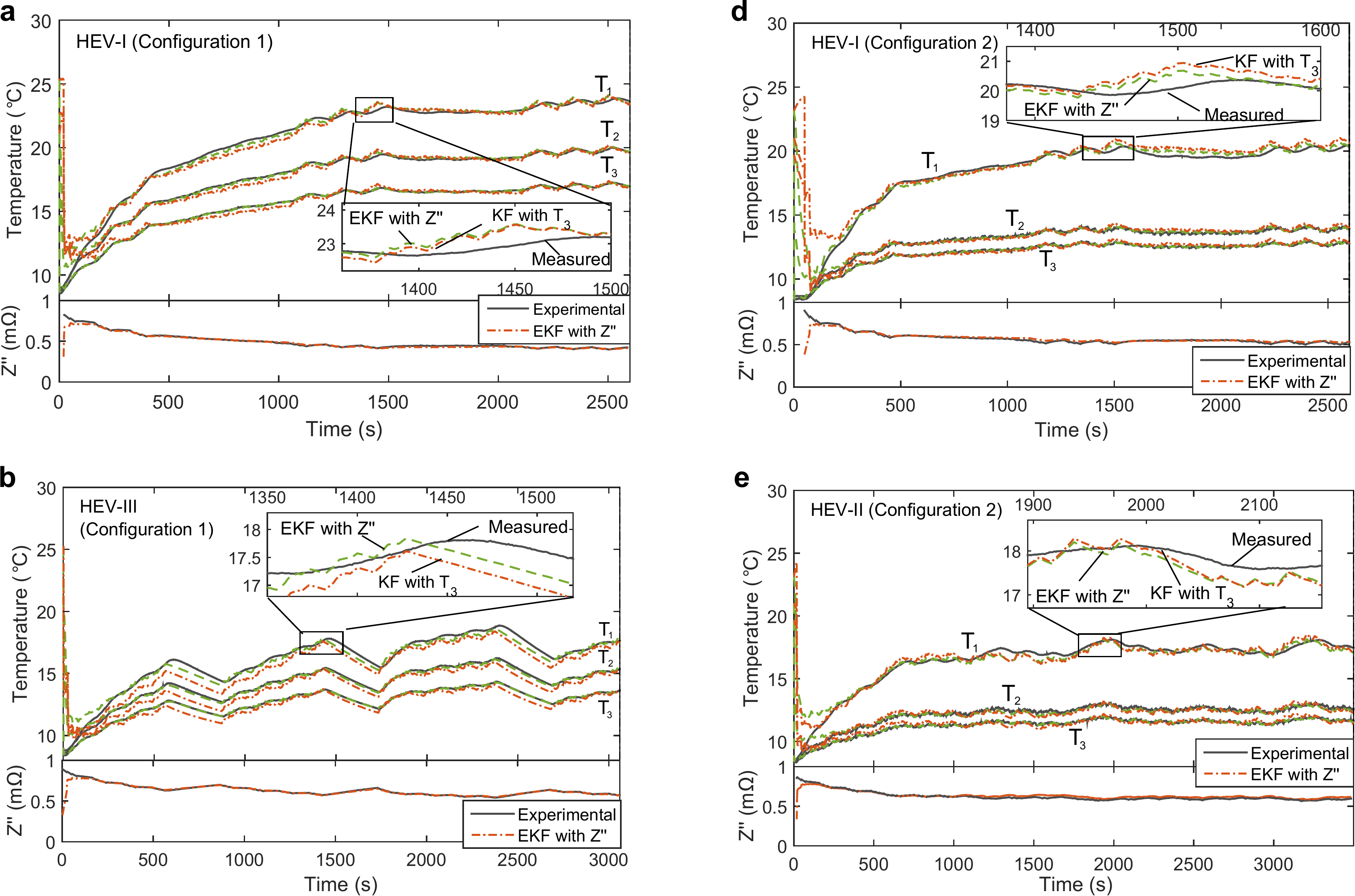}
		\includegraphics[width=0.89\textwidth]{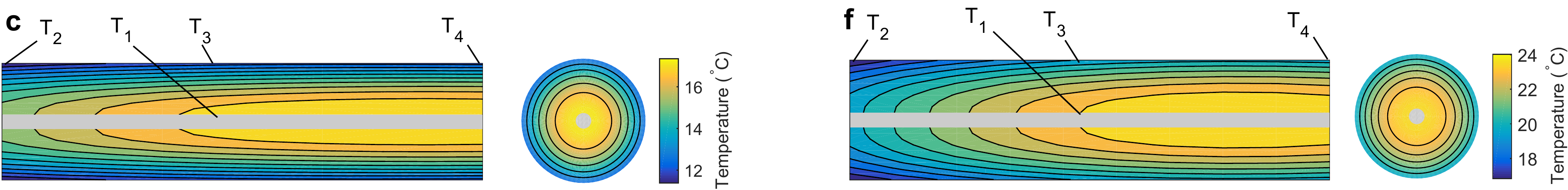}
		\caption{Comparison of temperature measurements against EKF (using $Z''$) and KF (using $T_{3}$). Config.\ 1: a) and b) time evolution of temperature outputs for HEV-I and HEV-III respectively;  c) 2-D contour plot at the end of the HEV-I cycle (t = 2550 s). Config.\ 2: d) and e) time evolution of temperature outputs for HEV-I and HEV-II respectively;  f) 2-D contour plot at the end of the HEV-II cycle (t = 3440 s).}
		\label{fig:main_fig}
	\end{figure*}
	
	\begin{figure*}
		\centering
		\includegraphics[width=0.42\textwidth]{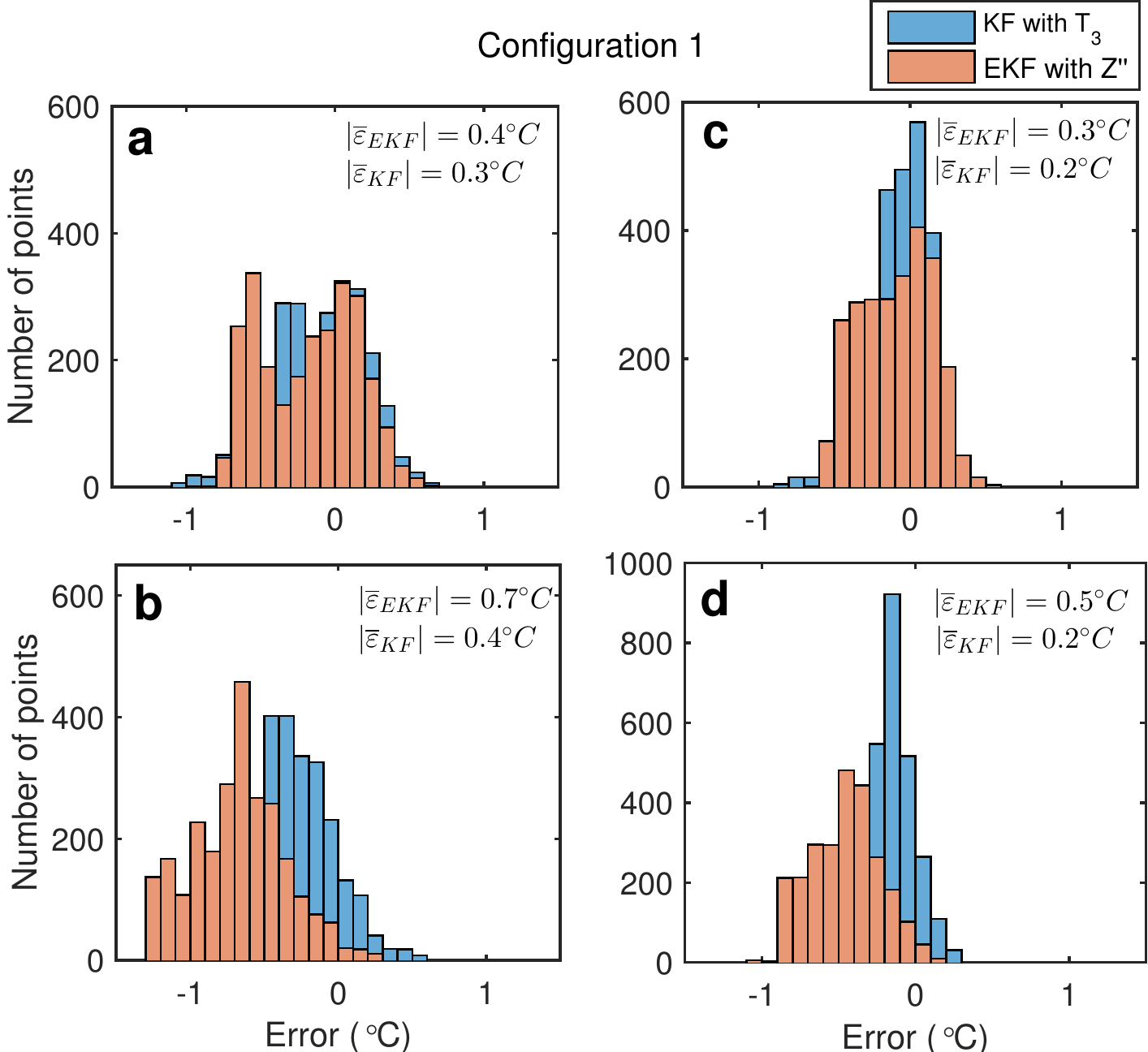}
		\hspace{0.9 cm}
		\includegraphics[width=0.42\textwidth]{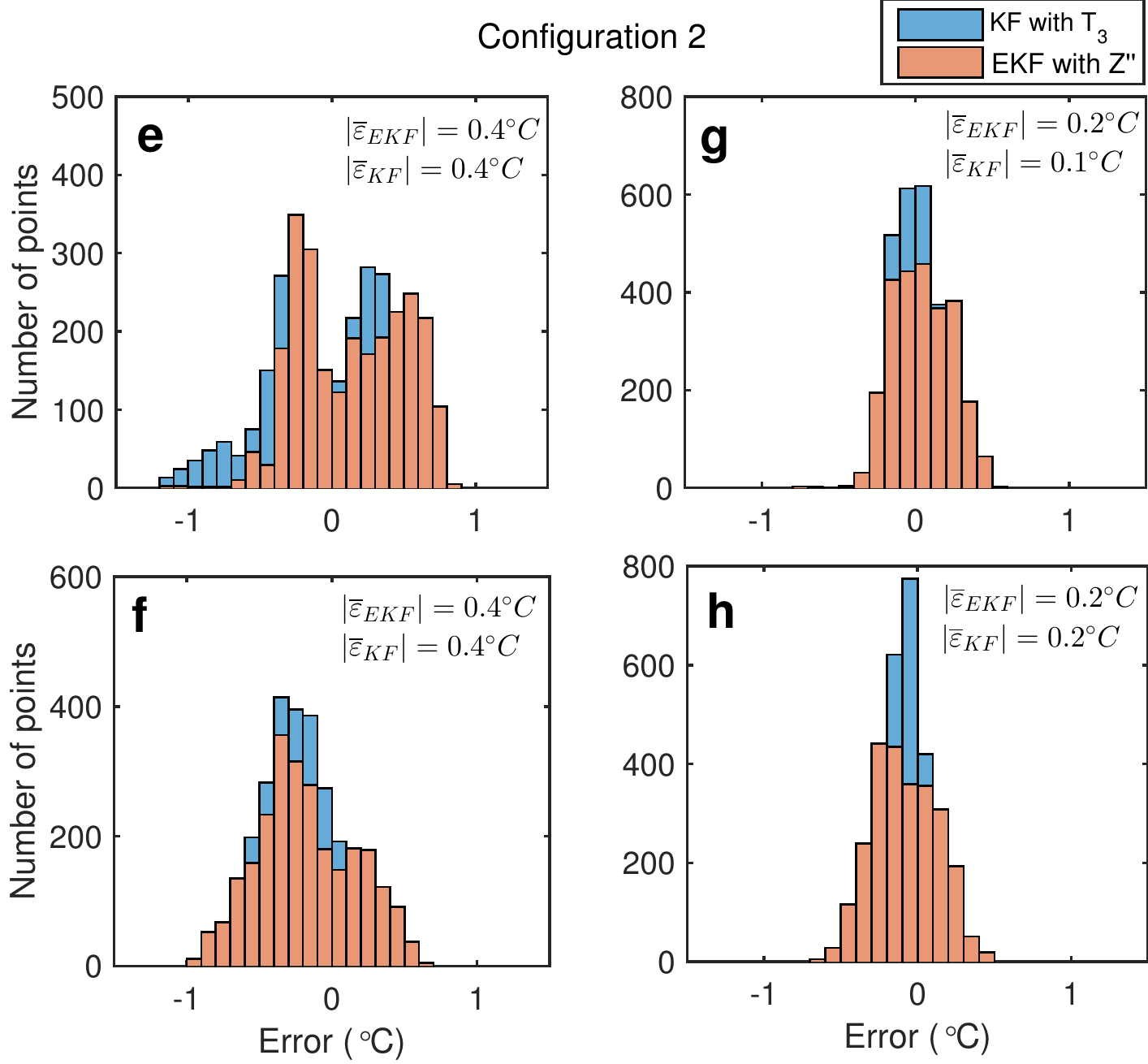}
		\caption{Error analysis showing the number of errors in the core temperature ($T_1$) or surface temperature ($T_3$) as denoted later in this caption, using the KF with surface temperature measurement (blue) and the EKF with $Z''$ measurement (orange). The overal RMSE value for each method is shown in the top corner of each plot. Configuration 1: a) HEV-I, $T_1$; b) HEV-III, $T_1$; c) HEV-I, $T_3$; d)  HEV-III, $T_3$.  Configuration 2: e) HEV-I, $T_1$; f)~HEV-II, $T_1$; g) HEV-I, $T_3$; h)  HEV-II, $T_3$.}
		\label{fig:hist_fig}
	\end{figure*}
}

We now compare the performance of the EKF with that of a KF based on the same thermal model but with $T_{3}$ as the measurement input rather than $Z''$.
{We chose $T_3$ for comparison since this is the location of the measurement input used in earlier studies~\cite{Kim2013,Kim2014b,richardson2015sensorless}.}
The initial state estimate provided to the battery is a uniform temperature distribution at 25 $^{\circ}$C, whereas the true initial battery state is a uniform temperature distribution at $8^{\circ}$C. The covariance matrices are calculated as $R^{n}=\sigma_{n}^{2}$ and $\mathbf{R}^{\mathbf{v}}=\beta_{v}^{2}diag(2,2)$.
The first tuning parameter for the EKF is chosen as $\sigma_{n}=3\times10^{-5}$ $\Omega$, based on the estimated standard deviation of the imaginary impedance measurement. The second tuning parameter was chosen as $\beta_{v}=5\times10^{-3}$, by trial and error. For the KF with $T_3$ as measurement input the tuning parameters were chosen as $\sigma_{n}=5\times10^{-4}$ $^\circ$C and $\sigma_{n}=0.05$, which are the same values as those used in our previous work~\cite{richardson2015sensorless}.

Figs.~\ref{fig:main_fig} a)-c) and d)-f) show the results for Config.\ 1 and Config.\ 2, respectively.  For each configuration, the core and surface temperatures quickly converge to the correct values and are accurately estimated throughout the rest of the excitation profile. Fig.~\ref{fig:main_fig}b shows that, for the HEV-III cycle, the EKF and KF underestimate the true temperature, in particular during periods with very low or zero applied currents. This error may be a result of a limitation in the thermal model: the heat generation term (eq.~(\ref{eq:Q})) only accounts for ohmic heating, whereas additional electrochemical heating may occur due to the relaxation of concentration gradients, which would result in continued internal heating after the removal of an applied load~\cite{bandhauer2011critical}. This error may be mitigated by the inclusion of an electrochemical heat generation model, although this is beyond the scope of the current work.
{Figs.~\ref{fig:main_fig}c) and f) show 2-D contour plots of the cell at the end of the HEV-I cycle and HEV-II cycle respectively. These contour plots are generated by the evaluating the cell temperature on a fine grid throughout the entire ($r$-$z$) domain using the thermal model described in Part I. They show that in each case, a significant radial temperature distribution develops within the cell, but in Configuration 2, a larger axial temperature gradient arises, as expected.}

Fig.~\ref{fig:hist_fig} shows histogram plots of the errors in the estimates of $T_1$ and $T_3$ in each drive cycle for the EKF with $Z''$ and the KF with $T_3$. Plots a)-d) apply to Config.\ 1 and plots e)-h) apply to Config.\ 2.
{Although the errors may include contributions from various sources (such as measurement uncertainty, variations in ambient temperature or fan flow rate, etc.), we can distinguish between impedance related and model related errors by considering the difference in error arising from the method using impedance measurement input versus surface temperature measurement input.}
In each case the errors are mostly peaked around a zero mean with a standard deviation in the range $0.2^\circ \text{C} - 0.7^\circ \text{C}$. The RMS error in each case is displayed in the top right corner of the plot.
It can be seen that the KF with $T_3$ performs slightly better in most cases, with a lower offset from the zero mean and a narrower distribution. This is to be expected given the higher accuracy of the thermocouple measurement. In particular, the KF performs better in the estimate of $T_3$, which is unsurprising since it uses $T_3$ as a measurement input. However, in general the performance of the EKF is satisfactory and comparable with that of the KF.
Fig.~\ref{fig:hist_fig}e shows that the error distribution for the estimate of $T_1$ for the HEV-III cycle is multi-modal: there are peaks at $\sim -0.2$ and $\sim 0.5 ^\circ \text{C}$. This is a manifestation of the errors discussed previously arising from the limitation of the purely ohmic heat generation term.

{We note that the RMSE of $T_1$ in Table~\ref{tab:param_validation_2d_table} is larger than that of the other three sensors ($T_2$, $T_3$ and $T_4$). This is perhaps explained as follows: (i) the parameterization scheme minimizes the error over all four sensors; but (ii) the error in each of the three surface temperatures may be strongly correlated since the temperature gradient in the radial direction is greater than that in the axial; and hence (iii) the parameterization may be biased towards minimizing the error in the three surface temperatures at the expense of the error in the core.}

{It should also be noted that since the uncertainty of the impedance measurement increases as impedance decreases, the temperature estimates become more uncertain at higher temperatures. Hence, the implementation of this technique could be more challenging at higher ambient temperature conditions than those studied here. This issue was addressed by Spinner et al.~\cite{spinner2015expanding}, who applied a secondary, empirical fit for the upper temperature range to improve accuracy. A similar approach could be applied using the hybrid method presented here, for applications involving higher temperature.}

\FloatBarrier
\section{Conclusions}

ITD can be used in conjunction with a thermal model to enable estimation of the spatially-resolved temperature distribution of lithium ion battery cells.
The extension of this method to 2-D conditions represents a significant improvement over earlier work, and opens up the possibility of applying this approach to a more general set of conditions than previously possible.
{For instance, the method could be used to monitor the internal temperature of cells in EV/HEV battery packs, which may have various different cooling configurations.}
Moreover, it could also be applied using alternative thermal models and/or battery geometries to provide robust estimation of the spatially resolved temperature field of batteries in a range of applications.
The calibration of the impedance coefficients using data from a single drive cycle is also a significant improvement over earlier methods. The reduced time and effort required for the calibration step makes practical implementation a feasible goal.

The accuracy of ITD is still slightly inferior to that of conventional methods based on surface thermocouples; however, the reduction in accuracy may be justified by the concurrent reduction in instrumentation cost/complexity. The application of this method online in a vehicle is perhaps the most important area of future work.
{A recent study has already demonstrated ITD (for average internal temperature estimation) in a vehicle~\cite{raijmakers2016non}. An interesting area for future work could involve applying the hybrid ITD approach presented here in an on-board setting.}

{An important consideration for the practical implementation of this approach is the independence of the impedance response to SoH. Whilst such independence has been verified for certain cell chemistries~\cite{Raijmakers2014d}, verification for each specific use case would be necessary.}
Moreover, the accuracy and precision of the impedance measurement will be a crucial factor in on-board applications, and devices capable of achieving high accuracy using low-cost equipment will {become increasingly} important.

\section*{Acknowledgements}
This work was funded by a NUI Travelling Scholarship, a UK EPSRC Doctoral Training Award,
the Foley-Bejar scholarship from Balliol College, University of Oxford, and the RCUK Energy Programmes's STABLE-NET project (ref. EP/L014343/1).

%

\section*{Appendix A}
{Matlab code for the model described in this paper is available online at www.github.com/robert-richardson/EKF-Impedance-2D-Temperature.}

\section*{References}

\bibliography{elsarticle-template}

\begin{thebibliography}{10}
\expandafter\ifx\csname url\endcsname\relax
  \def\url#1{\texttt{#1}}\fi
\expandafter\ifx\csname urlprefix\endcsname\relax\def\urlprefix{URL }\fi
\expandafter\ifx\csname href\endcsname\relax
  \def\href#1#2{#2} \def\path#1{#1}\fi

\bibitem{Forgez2010a}
C.~Forgez, D.~{Vinh Do}, G.~Friedrich, M.~Morcrette, C.~Delacourt,
  \href{http://linkinghub.elsevier.com/retrieve/pii/S037877530901982X}{{Thermal
  modeling of a cylindrical LiFePO4/graphite lithium-ion battery}}, Journal of
  Power Sources 195~(9) (2010) 2961--2968.
\newblock \href {http://dx.doi.org/10.1016/j.jpowsour.2009.10.105}
  {\path{doi:10.1016/j.jpowsour.2009.10.105}}.
\newline\urlprefix\url{http://linkinghub.elsevier.com/retrieve/pii/S037877530901982X}

\bibitem{Kim2013}
Y.~Kim, J.~B. Siegel, A.~G. Stefanopoulou,
  \href{http://ieeexplore.ieee.org/xpls/abs\_all.jsp?arnumber=6579917}{{A
  computationally efficient thermal model of cylindrical battery cells for the
  estimation of radially distributed temperatures}}, in: American Control
  Conference (ACC), 2013, Washington, DC,, 2013, pp. 698--703.
\newline\urlprefix\url{http://ieeexplore.ieee.org/xpls/abs\_all.jsp?arnumber=6579917}

\bibitem{Kim2014b}
Y.~Kim, S.~Mohan, S.~Member, J.~B. Siegel, A.~G. Stefanopoulou, Y.~Ding, {The
  Estimation of Temperature Distribution in Cylindrical Battery Cells Under
  Unknown Cooling Conditions}, IEEE Transactions on Control System Technology
  (2014) 1--10.

\bibitem{Lin2013f}
X.~Lin, H.~E. Perez, J.~B. Siegel, A.~G. Stefanopoulou, R.~D. Anderson, M.~P.
  Castanier, {Online Parameterization of Lumped Thermal Dynamics in Cylindrical
  Lithium Ion Batteries for Core Temperature Estimation and Health Monitoring},
  IEEE Trans. Control Syst. Technol. 21~(5) (2013) 1745--1755.
\newblock \href {http://dx.doi.org/10.1109/TCST.2012.2217143}
  {\path{doi:10.1109/TCST.2012.2217143}}.

\bibitem{Lin2014}
X.~Lin, H.~E. Perez, S.~Mohan, J.~B. Siegel, A.~G. Stefanopoulou, Y.~Ding,
  M.~P. Castanier,
  \href{http://linkinghub.elsevier.com/retrieve/pii/S0378775314001244}{{A
  lumped-parameter electro-thermal model for cylindrical batteries}}, Journal
  of Power Sources 257 (2014) 1--11.
\newblock \href {http://dx.doi.org/10.1016/j.jpowsour.2014.01.097}
  {\path{doi:10.1016/j.jpowsour.2014.01.097}}.
\newline\urlprefix\url{http://linkinghub.elsevier.com/retrieve/pii/S0378775314001244}

\bibitem{Pesaran2009}
A.~Pesaran, G.~H. Kim, M.~Keyser, {Integration Issues of Cells into Battery
  Packs for Plug-In and Hybrid Electric Vehicles}, in: EVS 24 International
  Battery, Hybrid and Fuel Cell Electric Vehicle Symposium, Stavanger, Norway,
  2009.

\bibitem{lin2014temperature}
X.~Lin, A.~G. Stefanopoulou, J.~B. Siegel, S.~Mohan, Temperature estimation in
  a battery string under frugal sensor allocation, in: ASME 2014 Dynamic
  Systems and Control Conference, American Society of Mechanical Engineers,
  2014, pp. V001T19A006--V001T19A006.

\bibitem{santhanagopalan2009analysis}
S.~Santhanagopalan, P.~Ramadass, J.~Z. Zhang, Analysis of internal
  short-circuit in a lithium ion cell, Journal of Power Sources 194~(1) (2009)
  550--557.

\bibitem{lee2011situ}
C.-Y. Lee, S.-J. Lee, M.-S. Tang, P.-C. Chen, In situ monitoring of temperature
  inside lithium-ion batteries by flexible micro temperature sensors, Sensors
  11~(10) (2011) 9942--9950.

\bibitem{mutyala2014situ}
M.~S.~K. Mutyala, J.~Zhao, J.~Li, H.~Pan, C.~Yuan, X.~Li, In-situ temperature
  measurement in lithium ion battery by transferable flexible thin film
  thermocouples, Journal of Power Sources 260 (2014) 43--49.

\bibitem{lee2015flexibleA}
C.-Y. Lee, H.-C. Peng, S.-J. Lee, I.~Hung, C.-T. Hsieh, C.-S. Chiou, Y.-M.
  Chang, Y.-P. Huang, et~al., A flexible three-in-one microsensor for real-time
  monitoring of internal temperature, voltage and current of lithium batteries,
  Sensors 15~(5) (2015) 11485--11498.

\bibitem{lee2015flexibleB}
C.-Y. Lee, S.-M. Chuang, S.-J. Lee, I.-M. Hung, C.-T. Hsieh, Y.-M. Chang, Y.-P.
  Huang, Flexible micro sensor for in-situ monitoring temperature and voltage
  of coin cells, Sensors and Actuators A: Physical 232 (2015) 214--222.

\bibitem{martiny2014development}
N.~Martiny, A.~Rheinfeld, J.~Geder, Y.~Wang, W.~Kraus, A.~Jossen, Development
  of an all kapton-based thin-film thermocouple matrix for in situ temperature
  measurement in a lithium ion pouch cell, Sensors Journal, IEEE 14~(10) (2014)
  3377--3384.

\bibitem{Srinivasan2011c}
R.~Srinivasan, B.~G. Carkhuff, M.~H. Butler, A.~C. Baisden, {Instantaneous
  measurement of the internal temperature in lithium-ion rechargeable cells},
  Electrochimica Acta 56~(17) (2011) 6198--6204.
\newblock \href {http://dx.doi.org/10.1016/j.electacta.2011.03.136}
  {\path{doi:10.1016/j.electacta.2011.03.136}}.

\bibitem{Srinivasan2012a}
R.~Srinivasan, {Monitoring dynamic thermal behavior of the carbon anode in a
  lithium-ion cell using a four-probe technique}, J. Power Sources 198 (2012)
  351--358.
\newblock \href {http://dx.doi.org/10.1016/j.jpowsour.2011.09.077}
  {\path{doi:10.1016/j.jpowsour.2011.09.077}}.

\bibitem{Schmidt2013a}
J.~P. Schmidt, S.~Arnold, A.~Loges, D.~Werner, T.~Wetzel, E.~Ivers-Tiff\'{e}e,
  {Measurement of the internal cell temperature via impedance: evaluation and
  application of a new method}, J. Power Sources 243 (2013) 110--117.
\newblock \href {http://dx.doi.org/10.1016/j.jpowsour.2013.06.013}
  {\path{doi:10.1016/j.jpowsour.2013.06.013}}.

\bibitem{Richardson2014}
R.~R. Richardson, P.~T. Ireland, D.~A. Howey,
  \href{http://linkinghub.elsevier.com/retrieve/pii/S0378775314006302}{{Battery
  internal temperature estimation by combined impedance and surface temperature
  measurement}}, Journal of Power Sources 265 (2014) 254--261.
\newblock \href {http://dx.doi.org/10.1016/j.jpowsour.2014.04.129}
  {\path{doi:10.1016/j.jpowsour.2014.04.129}}.
\newline\urlprefix\url{http://linkinghub.elsevier.com/retrieve/pii/S0378775314006302}

\bibitem{Raijmakers2014d}
L.~Raijmakers, D.~Danilov, J.~van Lammeren, M.~Lammers, P.~Notten, {Sensorless
  battery temperature measurements based on electrochemical impedance
  spectroscopy}, J. Power Sources 247 (2014) 539--544.
\newblock \href {http://dx.doi.org/10.1016/j.jpowsour.2013.09.005}
  {\path{doi:10.1016/j.jpowsour.2013.09.005}}.

\bibitem{richardson2015sensorless}
R.~R. Richardson, D.~A. Howey,
  \href{http://ieeexplore.ieee.org/xpls/abs_all.jsp?arnumber=7097077&tag=1}{{Sensorless
  Battery Internal Temperature Estimation using a Kalman Filter with Impedance
  Measurement}}, IEEE Transactions on Sustainable Energy 6~(4).
\newline\urlprefix\url{http://ieeexplore.ieee.org/xpls/abs_all.jsp?arnumber=7097077&tag=1}

\bibitem{Zhu2015}
J.~Zhu, Z.~Sun, X.~Wei, H.~Dai, {A new lithium-ion battery internal temperature
  on-line estimate method based on electrochemical impedance spectroscopy
  measurement}, Journal of Power Sources 274 (2015) 990--1004.
\newblock \href {http://dx.doi.org/10.1016/j.jpowsour.2014.10.182}
  {\path{doi:10.1016/j.jpowsour.2014.10.182}}.

\bibitem{spinner2015expanding}
N.~S. Spinner, C.~T. Love, S.~L. Rose-Pehrsson, S.~G. Tuttle, Expanding the
  operational limits of the single-point impedance diagnostic for internal
  temperature monitoring of lithium-ion batteries, Electrochimica Acta 174
  (2015) 488--493.

\bibitem{Koch2015}
R.~Koch, A.~Jossen, {Temperature Measurement of Large Format Pouch Cells with
  Impedance Spectroscopy}, EVS28 International Electric Vehicle Symposium and
  Exhibition (2015) 1--9.

\bibitem{Howey2014a}
D.~A. Howey, P.~D. Mitcheson, S.~Member, V.~Yufit, G.~J. Offer, N.~P. Brandon,
  {On-line measurement of battery impedance using motor controller excitation},
  IEEE Trans. Veh. Technol. 63~(6) (2014) 2557--2566.
\newblock \href {http://dx.doi.org/10.1109/TVT.2013.2293597}
  {\path{doi:10.1109/TVT.2013.2293597}}.

\bibitem{beelen2015improved}
H.~Beelen, L.~Raijmakers, M.~Donkers, P.~Notten, H.~Bergveld, An improved
  impedance-based temperature estimation method for li-ion batteries,
  IFAC-PapersOnLine 48~(15) (2015) 383--388.

\bibitem{troxler2014effect}
Y.~Troxler, B.~Wu, M.~Marinescu, V.~Yufit, Y.~Patel, A.~J. Marquis, N.~P.
  Brandon, G.~J. Offer, The effect of thermal gradients on the performance of
  lithium-ion batteries, Journal of Power Sources 247 (2014) 1018--1025.

\bibitem{richardson2016on}
R.~R. Richardson, S.~Zhao, D.~A. Howey, {On-board monitoring of 2-D
  spatially-resolved temperatures in cylindrical lithium-ion batteries: Part I.
  Low-order thermal modelling}, Arxiv Preprint (2016).

\bibitem{fleckenstein2013thermal}
M.~Fleckenstein, S.~Fischer, O.~Bohlen, B.~B{\"a}ker, Thermal impedance
  spectroscopy-a method for the thermal characterization of high power battery
  cells, Journal of Power Sources 223 (2013) 259--267.

\bibitem{hahn2012heat}
D.~W. Hahn, M.~N. Ozisik, Heat conduction, John Wiley \& Sons, 2012.

\bibitem{bandhauer2011critical}
T.~M. Bandhauer, S.~Garimella, T.~F. Fuller, A critical review of thermal
  issues in lithium-ion batteries, Journal of the Electrochemical Society
  158~(3) (2011) R1--R25.

\bibitem{Incropera2007a}
F.~P. Incropera, D.~P. {De Witt}, {Fundamentals of Heat and Mass Transfer}, 6th
  Edition, Wiley, 2007.

\bibitem{raijmakers2016non}
L.~H. Raijmakers, D.~L. Danilov, J.~P. van Lammeren, T.~J. Lammers, H.~J.
  Bergveld, P.~H. Notten, Non-zero intercept frequency: An accurate method to
  determine the integral temperature of li-ion batteries, IEEE Transactions on
  Industrial Electronics 63~(5) (2016) 3168--3178.

\end{thebibliography}

\end{document}